\documentclass[aps,pra,twocolumn,showpacs,superscriptaddress,letterpaper]{revtex4}

\usepackage{graphicx}
\usepackage{bm}
\usepackage{amsmath}
\usepackage{mathtools}
\usepackage{SIunits}
\usepackage{psfrag}
\usepackage{pstricks}
\usepackage{xspace}
\usepackage{color}
\usepackage[normalem]{ulem}
\usepackage{multirow}
\usepackage{ulem}

\newcommand{\ie}{i.e.\@\xspace}

\newcommand{\eq}[1]{Eq.~\eqref{#1}}
\newcommand{\eqs}[2]{Eqs.~\eqref{#1} and \eqref{#2}}
\newcommand{\eqss}[3]{Eqs.~\eqref{#1}, \eqref{#2} and \eqref{#3}}

\newcommand{\Eq}[1]{Equation~\eqref{#1}}
\newcommand{\fig}[1]{Fig.~\ref{#1}}

\newcommand{\tab}[1]{Table~\ref{#1}}

\renewcommand{\bm}[1]{\boldsymbol{\mathbf{#1}}} 
\newcommand{\ud}{\mathrm{d}}
\newcommand{\bra}{\left\langle}
\newcommand{\ket}{\right\rangle}

\newcommand{\hankel}{\operatorname{H}_0^{(1)}}
\newcommand{\bessel}{\operatorname{J}_0}
\newcommand{\keff}{k_{\text{eff}}}

\newcommand{\NoAutoSpaceBeforeFDP}[0]{}
\newcommand{\AutoSpaceBeforeFDP}[0]{}

\newcounter{tempa}
\newcounter{tempb}

\newenvironment{ddiag}[1]{\psset{unit=1.5mm,fillstyle=solid,fillcolor=white}
   \begin{pspicture}[shift=-6](0,-7)(#1,7)}{\end{pspicture}
}

\newcommand{\ggmoy}[3]{\psline[linewidth=0.5](#1,#3)(#2,#3)}
\newcommand{\eemoy}[3]{\psline[linewidth=0.5,linestyle=dashed](#1,#3)(#2,#3)}

\newcommand{\gggmoy}[4]{\psline[linewidth=0.5](#1,#2)(#3,#4)}
\newcommand{\eeemoy}[4]{\psline[linewidth=0.5,linestyle=dashed](#1,#2)(#3,#4)}
\newcommand{\pparticule}[2]{\pscircle(#1,#2){1}}
\newcommand{\iidentique}[4]{
   \psline(#1,#2)(#3,#4)
}

\newenvironment{dddiag}[1]{\psset{unit=1.5mm,fillstyle=solid,fillcolor=white}
   \begin{pspicture}[shift=-9](0,-10)(#1,10)}{\end{pspicture}
}
\newcommand{\hikami}[4]{
   \psframe[fillstyle=vlines,dimen=middle](#1,#2)(#3,#4)
   \setcounter{tempa}{#3}
   \addtocounter{tempa}{#1}
   \divide \value{tempa} by 2
   \setcounter{tempb}{#4}
   \addtocounter{tempb}{#2}
   \divide \value{tempb} by 2
   \pscircle(\value{tempa},\value{tempb}){2}
   \rput[c](\value{tempa},\value{tempb}){H}
}
\newcommand{\ladder}[4]{
   \psframe[fillstyle=none,dimen=middle](#1,#2)(#3,#4)
   \setcounter{tempa}{#3}
   \addtocounter{tempa}{#1}
   \divide \value{tempa} by 2
   \setcounter{tempb}{#4}
   \addtocounter{tempb}{#2}
   \divide \value{tempb} by 2
   \pscircle(\value{tempa},\value{tempb}){2}
   \rput[c](\value{tempa},\value{tempb}){L}
}

\begin{document}

\title{Intensity correlations between reflected and transmitted speckle patterns}

\author{N. Fayard}
\address{ESPCI ParisTech, PSL Research University, CNRS, Institut Langevin, 1 rue Jussieu, F-75005, Paris, France}
\author{A. Caz\'e}
\affiliation{Department of Mathematics, University of Michigan, Ann Arbor, Michigan 48109, USA}
\author{R. Pierrat}
\address{ESPCI ParisTech, PSL Research University, CNRS, Institut Langevin, 1 rue Jussieu, F-75005, Paris, France}
\author{R. Carminati}\email{remi.carminati@espci.fr}
\address{ESPCI ParisTech, PSL Research University, CNRS, Institut Langevin, 1 rue Jussieu, F-75005, Paris, France}
\date{\today}

\begin{abstract}
   We study theoretically the spatial correlations between the intensities measured at the input and output planes of a
   disordered scattering medium. We show that at large optical thicknesses, a long-range spatial correlation persists
   and takes negative values. For small optical thicknesses, short-range and long-range correlations coexist, with
   relative weights that depend on the optical thickness.  These results may have direct implications for
   the control of wave transmission through complex media by wavefront shaping, thus finding applications in sensing,
   imaging and information transfer.
\end{abstract}


\maketitle

\section{Introduction}

The study of wave scattering in disordered media is an active field of research, stimulated both
by innovative applications in imaging and sensing~\cite{SEBBAH-2001} and by fundamental questions 
in mesoscopic physics~\cite{SHENG-1995}. In the last few years, the possibility to control the propagation
of optical waves in complex media in the multiple scattering regime has been demonstrated using wavefront shaping 
techniques~\cite{VELLEKOOP-2007,CARMINATI-2010-4}. This breakthrough offers new
perspectives for imaging and communication through complex media~\cite{MOSK-2010,GIGAN-2010,FINK-2012}.
The initial schemes make use of optimization techniques requiring intensity measurements in the transmitted speckle,
which in terms of practical applications is a serious drawback. Finding a way to control the transmission and focusing
of light through a strongly scattering medium from measurements of the reflected speckle only is an issue of tremendous importance. 
Progresses have been made recently by taking advantage of the memory effect~\cite{VELLEKOOP-2010,MOSK-2012a,KATZ-2014-1}, 
with imaging capabilities limited to relatively small optical thicknesses. Nevertheless, the connection between 
the reflected and the transmitted speckle patterns generated by a disordered medium in the multiple scattering regime
has not been addressed theoretically so far.

In this paper, we make a step in this direction by studying theoretically and numerically the statistical correlation between the 
intensities measured in the transmitted and the reflected speckle patterns. The spatial intensity correlation function 
$C(\bm{r},\bm{r}^\prime)$ is defined as
\begin{equation}\label{correlation_def}
   C(\bm{r},\bm{r}^\prime)=\frac{\bra \delta I(\bm{r})\delta I(\bm{r}^\prime) \ket}{\bra I(\bm{r}) \ket \bra I(\bm{r}^\prime)\ket}
\end{equation}
where the notation $\bra\ldots\ket$ denotes a statistical average over disorder and $\delta I(\bm{r})= I(\bm{r})-\bra
I(\bm{r})\ket$ is the intensity fluctuation. This correlation function has been extensively studied in the context of
wave scattering and mesoscopic physics~\cite{BERKOVITS-1994,ROSSUM-1999,TIGGELEN-2003,MONTAMBAUX-2007}.  Theoretical
approaches often make use of the canonical slab or waveguide geometries (for a review
see~\cite{BERKOVITS-1994,ROSSUM-1999, BEENAKKER-1997-2} and references therein), where either transmitted or reflected
intensity is considered, or consider point sources in an infinite or open medium and compute intensity correlations for
two points inside or outside the medium~\cite{SHAPIRO-1986,SHAPIRO-1999,CAZE-2010,SARMA-2014,CARMINATI-2015-1}.  It
seems that the intensity correlation function for two points lying on different sides of a slab medium has not been
studied, and that the existence of a correlation has only been mentionned through passing~\cite{SAENZ-2003,FROUFE-2007-1}. In this
work, we study the correlation between the intensities in the input and output planes of a strongly scattering slab, as
sketched in Fig.~\ref{schema}. Using numerical simulations and analytical calculations, we show that for optically thick slabs a
correlation persists and takes negative values. Moreover, at smaller optical thicknesses, short and long-range correlations
coexist, with relative weights that depend on the optical thickness.  We believe these results to be a step
forward for the control of transmission through strongly scattering media, thus finding applications in sensing, imaging
and information transfer.
 
The spatial intensity correlation in a speckle pattern can be split into three contributions, historically denoted
by $C_1$, $C_2$ and $C_3$~\cite{FENG-1988,BERKOVITS-1994}, as follows:
\begin{equation}\label{correlation_decomposition}
   C(\bm{r},\bm{r}')=C_1(\bm{r},\bm{r}')+C_2(\bm{r},\bm{r}')+C_3(\bm{r},\bm{r}').
\end{equation}
The first term $C_1$ corresponds to the Gaussian statistics approximation for the field amplitude, and is a short-range
contribution, whose width determines the average size of a speckle spot~\cite{SHAPIRO-1986}. $C_2$ and $C_3$ are
non-Gaussian long-range correlations that decay on much larger scales~\cite{STEPHEN-1987,FENG-1988}.  In the diffusive
regime, and for two observation points lying on the same side of the scattering medium, these three contributions have different weights
such that $C_1\gg C_2\gg C_3$.  In the reflection/transmission configuration considered here, and at large optical thickness,
we show that an intensity correlation persists and is dominated by $C_2$ because of the short-range behavior of $C_1$. In
this case, the long-range character of $C_2$ is confered by its algebraic decay with respect to the distance between the two observation points.  
Moreover, this correlation is negative, a result that may have implications in the context of wave control by wavefront shaping.  For smaller
optical thicknesses, we also show that a crossover can be found between regimes dominated by $C_1$ and $C_2$,
respectively.

The paper is organized as follows. In Section~\ref{correlation_large_b}, we study the
reflection/transmission correlation function at large optical thicknesses. First, we use numerical simulations to compute
the correlation function without approximation and to describe its main features.
Second, we present the analytical calculation of the $C_2$ contribution to the
reflection/transmission correlation function in the multiple scattering regime, for both two-dimensional and three-dimensional
geometries, and show that $C_2$ is the leading contribution at large optical thicknesses. In Section~\ref{correlation_small_b}, 
we study the correlation function at small optical thicknesses, a regime in which the $C_1$ contribution dominates. 
In Section~\ref{stats}, as a consequence of the negative value of the correlation
that is found in the multiple scattering regime, we describe some peculiarities of the statistical distribution of
reflected/transmitted intensities. Finally, in Section~\ref{conclusion} we summarize the main results and discuss some
implications for the control of wave transmission through disordered media.

\section{Reflection/transmission correlation at large optical thickness}\label{correlation_large_b}

\subsection{Numerical analysis}

\begin{figure}[!htbf]
   \centering
   \psfrag{D}[c]{$D$}
   \psfrag{L}[c]{$L$}
   \psfrag{d}[c]{$\Delta r$}
   \psfrag{rr}[c]{$\bm{r}_R$}
   \psfrag{rt}[c]{$\bm{r}_T$}
   \psfrag{R}[c]{$R$}
   \includegraphics[width=0.8\linewidth]{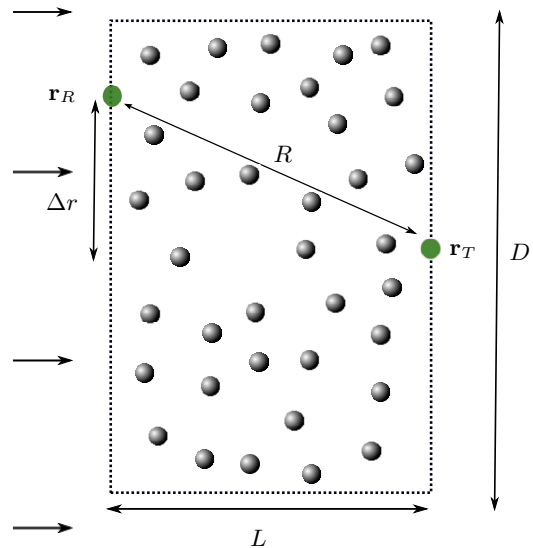} 
   \caption{ (Color online) Schematic representation of the scattering medium. The slab of thickness $L$ and transverse size $D$ is
   illuminated from the left by a monochromatic plane-wave at normal incidence. The correlation between the reflected
   and transmitted speckle patterns is characterized by the correlation function between the intensities
   at points $\bm{r}_R$ (reflection) and $\bm{r}_T$ (transmission).}
    \label{schema}
\end{figure}

In this section we present exact numerical simulations of wave scattering in the multiple scattering regime. We restrict ourselves to a
2D geometry for the sake of computer memory and time.  We consider a slab of scattering material, characterized by its
thickness $L$ and its transverse size $D$ (we keep $D>6 L$ in order to avoid finite size effects), as depicted in
\fig{schema}. Our purpose is the calculation of the correlation function $C(\bm{r}_R,\bm{r}_T)$, where $\bm{r}_R$ is a
point located on the input surface (reflection) and $\bm{r}_T$ is a point located on the output surface (transmission).

\subsubsection{Method}

To proceed, we use the coupled dipoles method~\cite{LAX-1952} to calculate numerically the intensity in the transmitted
and reflected speckle patterns. Repeating the calculations for a large number of configurations of disorder (positions
of scatterers) allows us to compute statistics.  The system contains $N$ randomly distributed non-overlapping point
scatterers, and is illuminated by a plane wave from the left at normal incidence. We deal with TE-polarized waves with
an electric field oriented along the invariance axis of the system (scalar waves).  The resonant point scatterers are
described by their electric polarizability
\begin{equation}\label{polarisability}
   \alpha(\omega)=-\frac{2\Gamma}{k_0^2(\omega-\omega_0+i\Gamma/2)}
\end{equation}
where $\omega_0$ is the resonance frequency and $\Gamma$ the linewidth. This specific form of the polarizability 
fulfils the optical theorem (\ie energy conservation). From the polarizability the
scattering cross section $\sigma(\omega)=k_0^3|\alpha(\omega)|^2/4$ and the scattering mean-free path
$\ell(\omega)=[\rho\sigma(\omega)]^{-1}$  can be deduced, where $\rho=N/(LD)$ is the number density of scatterers.
In the following, we consider scatterers at resonance ($\omega=\omega_0$) in order to reach large optical thicknesses
with a reasonable number of scatterers (typically a few hundreds).
In the coupled dipoles formalism, the exciting field $E_j$ on scatterer number $j$ is written
as~\cite{LAX-1952}
\begin{equation}\label{linear_system}
   E_j=E_0\left(\bm{r}_j\right)
   +\alpha\left(\omega\right)k_0^2\sum_{\substack{k=1\\k\ne j}}^NG_0\left(\bm{r}_j-\bm{r}_k\right)E_k
\end{equation}
where $G_0$ is the 2D free-space Green function given by $G_0(\bm{r}-\bm{r}')=(i/4)\hankel(k_0|\bm{r}-\bm{r}'|)$,
$\hankel$ being the Hankel function of first kind and order zero.
Equation (\ref{linear_system}) defines a set of $N$ linear equations that are solved
by a standard matrix inversion procedure. Once the exciting field is known on each scatterer, the field $E(\bm{r})$ 
and the intensity $I(\bm{r})=|E(\bm{r})|^2$ at any position
$\bm{r}$ inside or outside the scattering medium can be calculated by a direct summation, using
\begin{equation} \label{eq_E_total}
   E(\bm{r})=E_0\left(\bm{r}\right)
   +\alpha\left(\omega\right)k_0^2\sum_{j=1}^NG_0\left(\bm{r}-\bm{r}_j\right)E_j .
\end{equation}

\subsubsection{Numerical experiment}

\begin{figure}[!htbf]
   \centering
   \psfrag{x}[c]{$\Delta r/L$}
   \psfrag{y}[c]{$C(\Delta r)$}
   \psfrag{a}[l]{$C_2(\Delta r)$}
   \psfrag{b}[l]{$C_{\text{num}}(\Delta r)$}
   \includegraphics[width=1\linewidth]{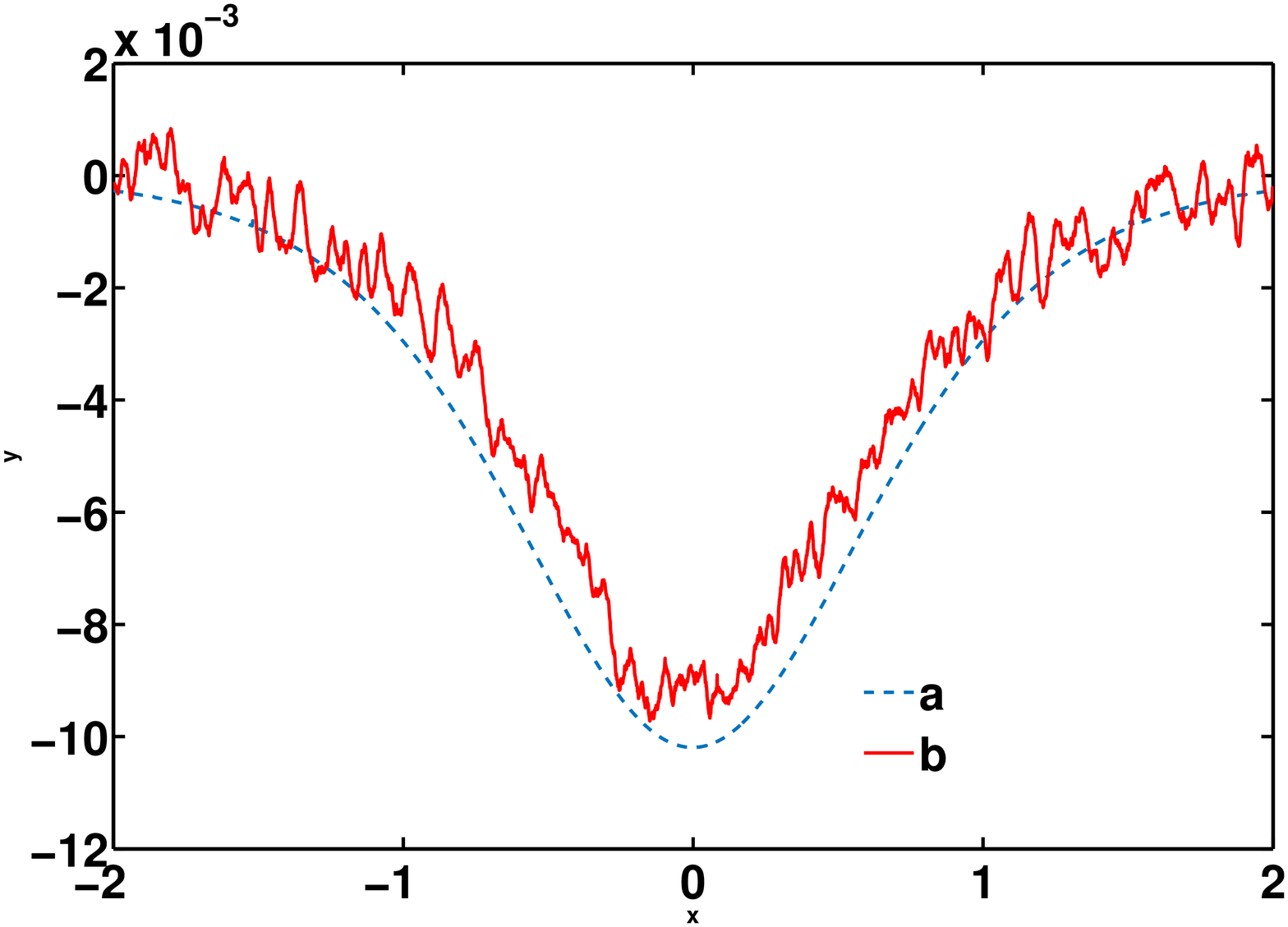}
   \caption{(Color online) Reflection/transmission intensity correlation $C_{\text{num}}$ given by the numerical simulations (red
   solid line) and analytical correlation $C_2$ given by \eq{C_2_final} (blue dashed line). Multiple scattering regime
   with $b=7$ and $k_0\ell =10$.}
   \label{C_2_large_b}
\end{figure}

We have carried out numerical simulations in the multiple scattering regime with an optical thickness $b=L/\ell=7$. 
This choice of optical thickness is
limited by the number of configurations that can be calculated in a reasonable computer time in order to get a
sufficiently accurate statistics to compute averaged values (requiring typically $10^6$ configurations). The correlation function
$C_{\text{num}}$ obtained from numerical calculation is plotted in \fig{C_2_large_b} (red solid line) versus the lateral shift
$\Delta r$ between the observation points in the reflected and transmitted speckles (see the geometry in \fig{schema}).

Surprisingly an intensity correlation subsists even for large optical thicknesses (multiple scattering regime). 
Moreover the reflection/transmission correlation function at large optical thickness takes a negative value around $\Delta r=0$.  
This means that $\bra\delta I(\bm{r}_R)\delta I(\bm{r}_T)\ket<0$, showing that the probability to have a dark spot in the transmitted
speckle in lateral coincidence with a bright spot in the reflection speckle (and {\it vice-versa}) should be substantial. This
property, that might have implications for the control of wave transmission by wavefront shaping, is investigated more
precisely in section~\ref{stats}.

\subsection{Analytical derivation of the $C_2$ reflection/transmission correlation function}

To get more insight on the numerical result presented above, we present the calculation of the $C_2$
contribution to the intensity correlation function for scalar waves in both two-dimensional (2D) and three-dimensional
(3D) geometries. Intuitively, we expected $C_2$ to be the leading contribution at large optical thicknesses, since $C_3$ is always 
negligible compared to $C_2$ and $C_1$ vanishes exponentially with the optical thickness in this particular reflection/transmission 
configuration.

For a dilute system such that $k_0 \ell \gg 1$, where $k=\omega/c=2\pi/\lambda$ with $c$ the speed of light and $\lambda$ 
the wavelength in vacuum, the intensity correlation functions can be calculated analytically using a perturbative approach.
Since the calculation for reflection-reflection or transmission-transmission correlations is detailed in 
textbooks or review articles~\cite{BERKOVITS-1994,SHENG-1995,MONTAMBAUX-2007}, we do not give all details
here but rather focus on the specificity of the reflection-transmission geometry. For the analytical derivation, we
consider that the transverse size $D$ of the slab is infinite.

\subsubsection{Average intensity}

To compute the correlation function $C_2$, we first need to compute the average intensity. The starting
point is the Bethe-Salpeter equation that reads~\cite{SHENG-1995,MONTAMBAUX-2007}
\begin{multline}\label{bethe-salpeter}
   \bra E(\bm{r},\omega)E^*(\bm{r'},\omega)\ket =  \bra E(\bm{r},\omega)\ket \bra E^*(\bm{r'},\omega)\ket
\\
   +\int \bra G(\bm{r},\bm{r}_{1},\omega)\ket \bra G^*(\bm{r'},\bm{r}_{2},\omega)\ket
\\
    \times \Gamma(\bm{r}_{1},\bm{r}_{2},\bm{r}_{3},\bm{r}_{4},\omega)\bra E(\bm{r}_{3},\omega)E^*(\bm{r}_{4},\omega)\ket  \ud\bm{r}_{1} \ud\bm{r}_{2} \ud\bm{r}_{3} \ud\bm{r}_{4} .
\end{multline}
In this equation, $\bra G(\bm{r},\bm{r}',\omega)\ket$ is the average Green function that links the average field $\bra
E(\bm{r},\omega)\ket$ to a source dipole $p$ located at position $\bm{r}'$ via
$\bra E(\bm{r},\omega)\ket=\mu_0\omega^2 \bra G(\bm{r},\bm{r}',\omega)\ket p.$
The operator $\Gamma(\bm{r}_{1},\bm{r}_{2},\bm{r}_{3},\bm{r}_{4},\omega)$ is the irreducible vertex that contains
all multiple scattering sequences connecting ($\bm{r}_3$, $\bm{r}_4$) to ($\bm{r}_1$, $\bm{r}_2$). 
The exact calculation of this complex object is out of reach, but an approximate expression to first-order in the small
parameter $1/(k_0\ell)$ and for independent scattering, known as the ladder approximation, can be derived~\cite{MONTAMBAUX-2007}.
Omitting the frequency dependence for simplicity in the following, it reads
\begin{equation}\label{vertex}
   \Gamma (\bm{r}_{1},\bm{r}_{2},\bm{r}_{3},\bm{r}_{4})=\gamma\delta(\bm{r}_1-\bm{r}_2)\delta(\bm{r}_3-\bm{r}_4)\delta(\bm{r}_1-\bm{r}_3)
\end{equation}
where
\begin{equation}\label{gamma}
   \gamma=
   \begin{cases}
      4k_0/\ell & \text{for 2D TE waves (2D),}
   \\
      4\pi/\ell & \text{for 3D scalar waves (3D).}
   \end{cases}
\end{equation}
Plugging Eq.~(\ref{vertex}) into Eq.~(\ref{bethe-salpeter}) yields
\begin{equation}\label{bethe-salpeter2}
   \bra I(\bm{r})\ket=\underbrace{\lvert\bra E(\bm{r})\ket\rvert^2}_{I_B(\bm{r})}
      +\underbrace{\gamma\int\lvert \bra G(\bm{r},\bm{r}')\ket \rvert^2 \bra I(\bm{r}')\ket\ud\bm{r}'}_{I_D(\bm{r})}.
\end{equation}
Physically, this closed equation means that only contributions for which $E$ and $E^*$ follow the same scattering path
have a significant weight in the average intensity (all cross terms vanish).  The first term $I_B$ in
\eq{bethe-salpeter2} is the ballistic (or coherent) component of the average intensity and the second term $I_D$ is the
diffuse part.  In order to get explicit expressions for these two quantities, we need to compute the average Green
function, or equivalently the average field.  As a consequence of the Dyson equation~\cite{MONTAMBAUX-2007}, in the
independent scattering limit, the average Green function obeys a propagation equation in an effective homogeneous
medium, defined by an effective wavevector $\keff$. We have
\begin{align}
   \bra G(\bm{r})\ket & = 
   \begin{dcases}
      \frac{i}{4}\hankel(\keff\lvert\bm{r}\rvert) & \text{(2D),}
   \\
      \frac{\exp(i\keff\lvert\bm{r}\rvert)}{4\pi\lvert\bm{r}\rvert} & \text{(3D),}
   \end{dcases}
\\
   \text{and} \bra E(\bm{r})\ket & = E_0\exp(i\keff z)
\end{align}
where $\keff =k_0+i/(2\ell)$ to first order in $1/(k_0\ell)$ and $z$ is the coordinate along the direction
normal to the slab. The ballistic intensity is readily
deduced:
\begin{equation}\label{ballistic}
   I_B(z)=I_0\exp[-z/\ell].
\end{equation}

Regarding the diffuse intensity $I_D(\bm{r})$, we can rewrite \eq{bethe-salpeter2} in the following way
\begin{equation}\label{bethe-salpeter3}
   I_D(\bm{r})=\gamma\int\lvert \bra G(\bm{r},\bm{r}')\ket \rvert^2 \left[I_B(\bm{r}')+I_D(\bm{r}')\right]\ud\bm{r}'.
\end{equation}
An analytical expression of the diffuse intensity can be obtained in the diffusive limit where
$|\bm{r}-\bm{r}'| \gg \ell$.  Making use of the translational invariance of the medium, the Fourier transform of
\eq{bethe-salpeter3} reads
\begin{equation}\label{bethe-salpeter4}
   I_D(\bm{q})=\gamma A(\bm{q})\left[I_B(\bm{q})+I_D(\bm{q})\right]
\end{equation}
where
\begin{equation}
   A(\bm{q})=\frac{1}{\gamma}\times\begin{dcases}
      \frac{1}{\sqrt{1+q^2\ell^2}} & \text{(2D),}
   \\
      \frac{\arctan(q\ell)}{q\ell} & \text{(3D).}
   \end{dcases}
\end{equation}
As we consider large distances, we can perform a second order Taylor expansion of $1/A(\bm{q})$ for $q\ell\ll 1$ which
leads in the real space to a diffusion-type equation
\begin{equation}\label{diffusion}
   -\frac{\ell^2}{d}\Delta I_D(\bm{r})=I_B(\bm{r})
\end{equation}
where $d\in\{2,3\}$ is the space dimension. Solving \eq{diffusion} in a slab geometry, we obtain
\begin{multline}
   I_D(z) = \frac{I_0 d}{L+2z_0}\left[(L+z_0-z)\left(1+\frac{z_0}{\ell}\right)\vphantom{\exp\left(-\frac{L}{\ell}\right)}\right.
\\
      \left.+(z+z_0)\left(1-\frac{z_0}{\ell}\right)\exp\left(-\frac{L}{\ell}\right)\right]
            -I_0 d\exp\left(-\frac{z}{\ell}\right)
\end{multline}
where $z_0$ is the extrapolation length needed to account for the boundary conditions at the input and exit surfaces of the
slab~\cite{CASE-1967,ISHIMARU-1997}. For a dilute index-matched slab, its expression is given
by~\cite{ISHIMARU-1997,LAI-2005}
\begin{equation}
   z_0=
   \begin{cases}
      \pi\ell/4 & \text{(2D),}
   \\
      2\ell/3 & \text{(3D).}
   \end{cases}
\end{equation}
This expression of the diffuse intensity $I_D$ is \emph{a priori} valid for distances $z$ such that $z \gg \ell$, for
which the diffusion approximation holds, but it surprisingly gives reasonably reliable results even for $z\le\ell$. Adding the
ballistic term, it also gives reliable results for the full average intensity
\begin{equation}\label{average_intensity}
   \bra I(z)\ket = I_B(z)+I_D(z)
\end{equation}
even for relatively small optical thicknesses.

\subsubsection{Long-range $C_2$ contribution}

The intensity correlation function is a fourth order correlation in terms of field amplitude.  Physically, a correlation
is created when the two pairs of fields that constitute the intensities in the correlation function share a common
history in the scattering process. Regarding $C_2$, the crossing occurs during the propagation of the intensities inside
the system and is described by a complex object known as a Hikami vertex~\cite{HIKAMI-1981}, and denoted by $H$ in the following.
The propagation of the intensity between the slab surfaces and the crossing is described by
the ladder operator, denoted by $L$ in the following. The expression of $C_2$ is given
by~\cite{HIKAMI-1981,BERKOVITS-1994,ROSSUM-1999,MONTAMBAUX-2007}
\begin{multline}\label{C_2}
   C_{2}(\bm{r}_R,\bm{r}_T)= \frac{1}{\bra I(\bm{r}_R)\ket \bra I(\bm{r}_T)\ket ]}\int \ud\bm{r}_{1}\ud\bm{r}_{2}\ud\bm{r}_{3}\ud\bm{r}_{4}
\\
   \times\ud\bm{\bm{\rho}}_{1} \ud\bm{\bm{\rho}}_{2} \ud\bm{\bm{\rho}}_{3} \ud\bm{\bm{\rho}}_{4} \lvert\bra G(\bm{r}_R,\bm{r}_{2})\ket \rvert^2\lvert \bra G(\bm{r}_T,\bm{r}_{4})\ket \rvert^2   
\\
   \times   L(\bm{r}_{2},\bm{\bm{\rho}}_{2})  L(\bm{r}_{4},\bm{\bm{\rho}}_{4})H(\bm{\bm{\rho}}_{1},\bm{\bm{\rho}}_{2},\bm{\bm{\rho}}_{3},\bm{\bm{\rho}}_{4}) 
\\
    \times  L(\bm{\bm{\rho}}_{1},\bm{r}_{1}) L(\bm{\bm{\rho}}_{3},\bm{r}_{3}) \lvert \bra E(\bm{r}_{1})\ket  \rvert^2 \lvert\bra E(\bm{r}_{3})\ket  \rvert^2
\end{multline}
which can be rewritten diagrammatically in the following form:
\begin{multline}\label{C_2_diag}
   C_2(\bm{r}_R,\bm{r}_T)=
\\
   \begin{dddiag}{51}
      \eemoy{0}{6}{-9}
      \eemoy{0}{6}{-3}
      \eemoy{0}{6}{9}
      \eemoy{0}{6}{3}
      \ladder{6}{-9}{18}{-3}
      \ladder{6}{3}{18}{9}
      \ladder{30}{-9}{42}{-3}
      \ladder{30}{3}{42}{9}
      \hikami{18}{-9}{30}{9}
      \rput[Bl](49,5){$\bm{r}_R$}
      \rput[Bl](49,-7){$\bm{r}_T$}
      \gggmoy{42}{-9}{48}{-6}
      \gggmoy{42}{9}{48}{6}
      \gggmoy{42}{-3}{48}{-6}
      \gggmoy{42}{3}{48}{6}
   \end{dddiag}
\end{multline}
where the hatched box is the Hikami vertex and the other boxes are the ladder operators. Thick lines represent the
average Green function and thick dashed lines stand for the average electric field.

The ladder operator is defined as
\begin{equation}\label{ladder}
   L(\bm{r},\bm{r}') = \gamma \delta(\bm{r}-\bm{r}')
      +\gamma\int\lvert\bra G(\bm{r},\bm{r}'')\ket\rvert^2 L(\bm{r}'',\bm{r}') \ud\bm{r}''
\end{equation}
and is represented diagrammatically as
\begin{equation}\label{ladder_diag}
   L(\bm{r},\bm{r}') =
   \begin{ddiag}{2}
      \iidentique{1}{3}{1}{-3}
      \pparticule{1}{-3}
      \pparticule{1}{3}
   \end{ddiag}+
   \begin{ddiag}{8}
      \ggmoy{1}{7}{-3}
      \ggmoy{1}{7}{3}
      \iidentique{1}{3}{1}{-3}
      \pparticule{1}{-3}
      \pparticule{1}{3}
      \iidentique{7}{3}{7}{-3}
      \pparticule{7}{-3}
      \pparticule{7}{3}
   \end{ddiag}+
   \begin{ddiag}{14}
      \ggmoy{1}{7}{-3}
      \ggmoy{1}{7}{3}
      \ggmoy{7}{13}{-3}
      \ggmoy{7}{13}{3}
      \iidentique{1}{3}{1}{-3}
      \iidentique{7}{3}{7}{-3}
      \iidentique{13}{3}{13}{-3}
      \pparticule{1}{-3}
      \pparticule{1}{3}
      \pparticule{7}{-3}
      \pparticule{7}{3}
      \pparticule{13}{-3}
      \pparticule{13}{3}
   \end{ddiag}+\ldots
\end{equation}
where circles and thick horizontal solid lines represent scattering events and average Green
functions, respectively, the top line standing for the field amplitude $E$ and the bottom line for its complex conjugate $E^*$. 
Thin vertical solid lines link scattering events involving identical scatterers.

An analytical expression of the ladder operator can be obtained in the diffusive limit previously used for
the computation of the average intensity. Making use of the translational invariance along the direction of the 
slab interface, the Fourier transform of the ladder operator with respect to transverse variables can be obtained
from \eq{ladder}, and reads
\begin{multline}\label{ladder2}
   L(\bm{K},z,z')= \frac{d \, \gamma}{\ell^2 K}
\\
   \times \frac{\sinh[K(z_<+z_0)]\sinh[K(L+z_0-z_>)]}{\sinh[K(L+2z_0)]}
\end{multline}
where $z_<=\min(z,z')$, $z_>=\max(z,z')$.

The Hikami box can also be calculated in the limit $k_0\ell\gg 1$, and its expression reduces to~\cite{BERKOVITS-1994,ROSSUM-1999,MONTAMBAUX-2007}
\begin{equation}\label{hikami}
   H(\bm{\rho}_{1},\bm{\rho}_{2},\bm{\rho}_{3},\bm{\rho}_{4})
      =h\int\prod_{i=1}^4\delta(\bm{\rho}-\bm{\rho}_{i})\bm{\nabla}_{\bm{\rho}_2}\cdot\bm{\nabla}_{\bm{\rho}_4}\ud\bm{\rho}
\end{equation}
where
\begin{equation}
   h=\begin{dcases}
      \ell^5/(16k_0^3) & \text{(2D),}
   \\
      \ell^5/(24\pi k_0^2) & \text{(3D).}
   \end{dcases}
\end{equation}
To have an explicit expression of $C_2$, we first compute the integrals involving $\bm{r}_1$ and $\bm{r}_3$ using 
\begin{equation}\label{ladder_average_intens}
   \int  L(\bm{r},\bm{r}')\lvert\bra E(\bm{r}')\ket\rvert^2\ud\bm{r}'=\gamma\bra I(\bm{r})\ket.
\end{equation}
As we deal with large optical thicknesses, we can replace the average intensity by its diffuse component.  Regarding the
integrals involving $\bm{r}_2$ and $\bm{r}_4$, we assume that the ladder operators vary slowly at the scale of the scattering
mean-free path $\ell$. This amounts to replacing $\bm{r}_2$ by $\bm{r}_R$ and $\bm{r}_4$ by $\bm{r}_T$ in the
ladder positions.  We end up with an explicit expression of the $C_2$ contribution to the correlation function, given by
\begin{multline}\label{C_2_final}
   C_2(\bm{r}_R,\bm{r}_T)= \frac{-I_0^2}{\bra I(\bm{r}_R)\ket \bra I(\bm{r}_T)\ket}
      \int_0^{+\infty}\frac{\ud q}{2q^2} F\left(\frac{q\Delta r}{L}\right)
\\
   \times\left[\left\{q^2\left(1+\frac{2z_0}{L}+\frac{2z_0^2}{L^2}\right)+1\right\}\sinh(q)-q\cosh(q)\right]
\\
   \times\frac{L^2}{(L+2z_0)^2}\frac{\sinh(qz_0/L)^2}{\sinh[q(1+2z_0/L)]^2}
\end{multline}
where
\begin{equation}
   F\left(\frac{q\Delta r}{L}\right)=
      \left[1+\frac{z_0}{\ell}\right]^2
   \times\begin{dcases}   
      \frac{16}{\pi k_0\ell}\frac{\cos(q\Delta r/L)}{q} & \text{(2D),}
   \\
      \frac{27}{k_0^2\ell L}\bessel(q\Delta r/L) & \text{(3D).}
   \end{dcases}
\end{equation}
The expression of the $C_2$ contribution to the reflection/transmission correlations function, together with its
comparison to full numerical simulations, is the main result of this paper. 
The dependence of the amplitude on the system size $L$ is approximatively $L^{1-d}$,
with $d$ the dimension of space. This shows that even at large optical thicknesses, a correlation subsists
between the intensities measured in the transmitted and the reflected speckles, and that this correlation is dominated
by a contribution of the $C_2$ type.

It is important to keep in mind that this expression has been derived in the framework of the diffusion approximation (in particular
bulk average Green functions have been used). Its validity in the geometry considered here where both reflected and
transmitted intensities contribute is checked by comparison to full numerical simulations in \fig{C_2_large_b}.  It is
interesting to note that the analytical and numerical calculations are in very good quantitative agreement, showing that the
diffusion approximation, used to derive the analytical results, is very accurate even for an optical thickness
$b=7$ (that is not very large) and a geometry involving a reflected intensity (that always involves short scattering paths).

We have seen that $C_2$ decays algebraically with the system size $L$. Thus it is interesting to analyse the behavior of the full
correlation function at smaller optical thicknesses, and in the crossover between the multiple and single scattering regimes.
This problem can be addressed numerically, by using the numerical method described previously
in various scattering regimes, and is the subject of the following section.

\section{Reflection/transmission correlation at small optical thickness}\label{correlation_small_b}

\subsection{Numerical analysis}

\begin{figure}[!htbf]
   \centering
   \psfrag{x}[c]{$\Delta r/L$}
   \psfrag{y}[c]{$C(\Delta r)$}
   \psfrag{a}[l]{$C_1+C_1'+C_2$}
   \psfrag{b}[l]{$C_{\text{num}}(\Delta r)$}
   \includegraphics[width=1\linewidth]{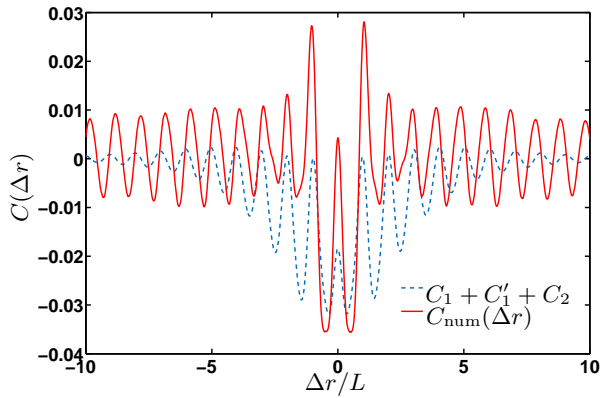}
   \caption{(Color online) Reflection/transmission intensity correlation $C_{\text{num}}$ given by the numerical simulations (red
   solid line) and analytical correlation including all terms $C_1+C_1'+C_2$ (blue dashed line). Single scattering regime
   with $b=0.5$ and $k_0\ell =10$.}
   \label{C_2_small_b}
\end{figure}

We consider a small optical thickness $b=0.5$, corresponding to the single-scattering regime. The reflection/transmission
intensity correlation function calculated numerically in this regime is shown in \fig{C_2_small_b} (red solid line). 
The large oscillations observed on a scale on the order of the wavelength are not described by the
$C_2$ contribution, and are expected to be a signature of the short-range $C_1$ contribution. To check the validity of this assumption, we have to
compute the $C_1$ contribution to the correlation function. At small optical thickness, a quantitative calculation would require
to go beyond the diffusion approximation and to account properly for the boundary conditions~\cite{FREUND-1992}. Since our purpose in
this section is only to support qualitatively the analysis of the general trends observed in the numerical simulations using a simple model,
we keep using the diffusion approximation to estimate the $C_1$ contribution.

\subsection{$C_1$ contribution and specificity of the reflection/transmission configuration}

In the scattering sequences picture, the $C_1$ correlation is created by interchanging the amplitudes between two independent ladders
at the last scattering event. The $C_1$ contribution reads
\begin{multline}\label{C_1}
   C_1(\bm{r}_R,\bm{r}_T)=\frac{1}{\bra I(\bm{r}_R)\ket \bra I(\bm{r}_T)\ket}
   \int \ud\bm{r}_{1}\ud\bm{r}_{2}\ud\bm{r}_{3}\ud\bm{r}_{4}  
\\  
   \times \bra G^*(\bm{r}_{R},\bm{r}_4)\ket
   \bra G(\bm{r}_{T},\bm{r}_4)\ket  \bra G(\bm{r}_{R},\bm{r}_2)\ket \bra G^*(\bm{r}_{T},\bm{r}_2)\ket
\\  
   \times L(\bm{r}_{2},\bm{r}_{1})L(\bm{r}_{4},\bm{r}_{3}) \lvert \bra E(\bm{r}_{1})\ket \rvert^2\lvert \bra E(\bm{r}_{3})\ket \rvert^2.
\end{multline}
In terms of diagrams, \eq{C_1} can be rewritten as
\begin{equation}\label{C_1_diag}
   C_1(\bm{r}_R,\bm{r}_T)=
   \begin{dddiag}{35}
      \rput[Bl](31,-1){$\bm{r}_R$}
      \rput[Bl](25,-1){$\bm{r}_T$}
      \ladder{6}{-9}{18}{-3}
      \ladder{6}{3}{18}{9}
      \gggmoy{18}{-9}{30}{0}
      \gggmoy{18}{9}{30}{0}
      \gggmoy{18}{-3}{24}{0}
      \gggmoy{18}{3}{24}{0}
      \eemoy{0}{6}{-9}
      \eemoy{0}{6}{-3}
      \eemoy{0}{6}{9}
      \eemoy{0}{6}{3}
   \end{dddiag}.
\end{equation}
In \eq{C_1}, the scattering of both pairs of fields from points $\bm{r}_1$ to $\bm{r}_2$ and from $\bm{r}_3$ to
$\bm{r}_4$, respectively, is described by a ladder propagator. The mixing of amplitudes at the last scattering event is
represented by four different average Green functions. The integrals in \eq{C_1} can be factorized, leading to
\begin{equation}\label{C_1_factorized}
   C_1(\bm{r}_R,\bm{r}_T)=\frac{\lvert B(\bm{r}_R,\bm{r}_T)\rvert^2}{\bra I(\bm{r}_R)\ket \bra I(\bm{r}_T)\ket}
\end{equation}
where
\begin{multline}\label{C_1_B}
   B(\bm{r}_R,\bm{r}_T)=\int
    \bra G(\bm{r}_{R},\bm{r}_2)\ket\bra G^*(\bm{r}_{T},\bm{r}_2)\ket 
\\
   \times L(\bm{r}_{2},\bm{r}_{1}) \lvert \bra E(\bm{r}_{1})\ket \rvert^2 \ud\bm{r}_{1}\ud\bm{r}_{2}.
\end{multline}
\Eq{C_1_factorized} shows that $C_1$ is the square of the scattered field correlation function. Indeed,
$C_1$ can also be seen as the correlation that would be observed for a field with Gaussian statistics, for which this
factorization holds~\cite{SEBBAH-2002-1}. Starting from \eqs{C_1_factorized}{C_1_B}, the usual way to derive the analytical
expression of $C_1$ consists in replacing $\bm{r}_2$ by $\bm{r}_T$ (for a speckle computed in transmission) or $\bm{r}_2$ by
$\bm{r}_R$ (for a speckle computed in reflection) in the ladder positions. In the reflection/transmission geometry,
this simplification cannot be performed because the relative distance between the points $\bm{r}_T$ and
$\bm{r}_R$ can be very large compared to the scattering mean-free path $\ell$. However, in order to get an explicit
expression, we can make use of \eq{ladder_average_intens} which leads to
\begin{multline}\label{C_1_B_2}
   B(\bm{r}_R,\bm{r}_T)=\int\gamma
   \bra G(\bm{r}_{R},\bm{r}_2)\ket
   \bra G^*(\bm{r}_{T},\bm{r}_2)\ket \bra I(\bm{r}_{2})\ket\ud\bm{r}_{2}.
\end{multline}
The integration over $\bm{r}_2$ is then performed using the residue theorem (the details are
given in appendix~\ref{C_1_calculation}). This leads to
\begin{multline}\label{C_1_B_3}
   B(\bm{r}_R,\bm{r}_T)=\gamma I_0\int\frac{\exp\left[-(ik'+k'')L-i\bm{K}\cdot\bm{\Delta r}\right]}{4(k'^2+k''^2)}
\\
      \times\left[\frac{d\left\{(1+z_0/\ell)M_1+(1-z_0/\ell)\exp(-L/\ell)M_2\right\}}{L+2z_0}\right.
\\
      \left.\vphantom{\frac{1}{2}}-(d-1)M_3\right]\frac{\ud\bm{K}}{(2\pi)^{d-1}}
\end{multline}
where $M_1$, $M_2$, $M_3$, $k'$ and $k''$ are given by \eqss{M_1}{M_2}{M_3} and \eq{kp_kpp}, respectively. The last
integral over $\bm{K}$ is performed numerically.

In the specific geometry considered here, another contribution has to be added, in which the ballistic intensity
contributes as one of the intensities involved in the correlation function~\cite{ROGOZKIN-1995}. Indeed, the ladder
operator involves at least one scattering event, and does not account for situations in which there is no scattering
event before the field interchange. Such contributions to the correlation function can be important for small optical thicknesses, where
the ballistic contribution is not negligible. This leads to a correction to the $C_1$ correlation function, that we
denote by $C_1'$, and whose expression is
\begin{multline}\label{C_1p}
   C_1'(\bm{r}_R,\bm{r}_T)= \frac{1}{\bra I(\bm{r}_R)\ket \bra I(\bm{r}_T)\ket}
      \bra E(\bm{r}_R) \ket \bra E^*(\bm{r}_T) \ket
\\
   \times\int 
      \bra G^*(\bm{r}_{R},\bm{r}_2)\ket \bra G(\bm{r}_{T},\bm{r}_2)\ket L(\bm{r}_{2},\bm{r}_{1}) \lvert \bra E(\bm{r}_{1})\ket \rvert^2   \ud\bm{r}_{1}\ud\bm{r}_{2}
\\
      + \text{c.c.}
\end{multline}
In terms of diagrams, the above expression can be rewritten as
\begin{equation}\label{C_1p_diag}
   C_1'(\bm{r}_R,\bm{r}_T)=
   \begin{dddiag}{33}
      \rput[Bl](31,-1){$\bm{r}_R$}
      \rput[Bl](25,-1){$\bm{r}_T$}
      \ladder{6}{-9}{18}{-3}
      \gggmoy{18}{-9}{30}{0}
      \eeemoy{18}{9}{30}{0}
      \gggmoy{18}{-3}{24}{0}
      \eeemoy{18}{3}{24}{0}
      \eemoy{0}{6}{-9}
      \eemoy{0}{6}{-3}
      \eemoy{0}{18}{9}
      \eemoy{0}{18}{3}
   \end{dddiag}
   +\text{c.c.}
\end{equation}
Making use of the quantity $B(\bm{r}_R,\bm{r}_T)$ defined in \eq{C_1_B}, we obtain
\begin{multline}\label{C_1p_factorized}
   C_1'(\bm{r}_R,\bm{r}_T)=\frac{2\Re\left[\bra E^*(\bm{r}_R)\ket\bra E(\bm{r}_T)\ket B(\bm{r}_R,\bm{r}_T)\right]}
      {\bra I(\bm{r}_R)\ket \bra I(\bm{r}_T)\ket}.
\end{multline}
It is important to note that the $C_1'$ contribution introduced here, and that involves the average field, is
not negligible compared to the usual $C_1$ contribution.  This correction should be added to the $C_1$ contribution when the
product of the ballistic fields $\bra E^*(\bm{r}_R)\ket\bra E(\bm{r}_T)\ket$ cannot be neglected, as
in the reflection/transmission geometry at low optical thickness, or in the reflection/reflection
geometry at any optical thickness.  We can show from \eq{C_1_B_3} that the $C_1$ and $C_1'$ contributions to the
correlation function decrease exponentially with the optical thickness $b=L/\ell$. This behavior explains why $C_2$
dominates at large optical thickness. But in the single scattering regime, an important contribution of the $C_1+C_1'$
term is observed. This is clearly seen in \fig{C_2_small_b} (blue dashed line), in which the sum of the three contributions $C_1+C_1'+C_2$
calculated analytically in the same two-dimensional geometry used for the numerical simulation is plotted.
Qualitatively, the behavior observed in the numerical simulation is fairly reproduced by the analytical approach. 

Moreover, since the $C_1+C_1'$ contribution decays exponentially with the medium thickness $L$, a crossover is expected towards
a regime dominated by $C_2$ when the optical thickness increases. This also shows that the reflection/transmission geometry studied here 
may be relevant to put forward experimentally the influence of the $C_2$ contribution 
(in the pure reflection or transmission geometries, the $C_1$ contribution is always the leading contribution).

\section{Statistical distribution of reflected and transmitted intensities}\label{stats}

Analytical and numerical results at large optical thickness have shown that a correlation between reflected
and transmitted intensities exists. Surprisingly, this correlation function takes negative values around $\Delta r=0$.
Having $\bra\delta I(\bm{r}_R)\delta I(\bm{r}_T)\ket<0$ for $\Delta r=0$ qualitatively suggests a high probability to have
a bright (dark) spot in the transmitted speckle in coincidence with a dark (bright) spot in the reflected speckle. 
In order to address this question in more quantitative terms, we have studied the full statistical
distribution of the intensities. More precisely, from the numerical simulations, we have extracted the statistical distributions 
of the product of the fluctuating part of the intensities at $\Delta r=0$, defined as
\begin{equation}
P(\tilde{\delta I_R}\tilde{\delta I_T}) \equiv P \left [  \frac{\delta I(\bm{r}_R)\delta
   I(\bm{r}_T)}{\bra I(\bm{r}_R)\ket\bra I(\bm{r}_T)\ket} \right ]
\end{equation}
whose average value is the intensity correlation $C(\bm{r}_R,\bm{r}_T)$ at $\Delta r=0$.

\begin{figure}[!htbf]
   \centering
   \psfrag{x}[c]{$\tilde{\delta I_R}\tilde{\delta I_T}$}
   \psfrag{y}[c]{$P(\tilde{\delta I_R}\tilde{\delta I_T})$}
   \includegraphics[width=1\linewidth]{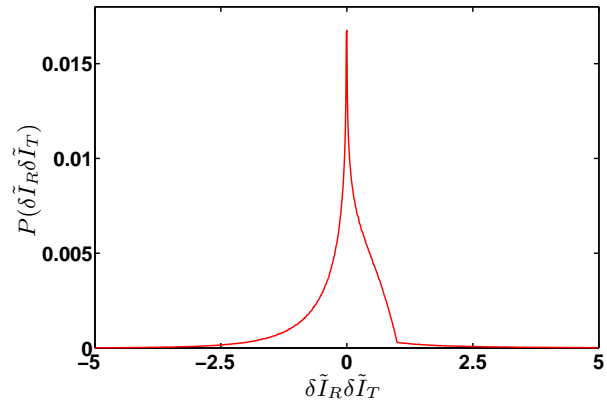}
   \caption{ (Color online) Statistical distribution of the product of the normalized reflected/transmistted intensities 
   $P(\tilde{\delta I_R}\tilde{\delta I_T})$ at $\Delta r=0$. Multiple scattering regime with $b=7$ and $k_0\ell=10$.}
   \label{histo}
\end{figure}

The statistical distribution $P(\tilde{\delta I_R}\tilde{\delta I_T})$ obtained for $b=7$ and $k_0\ell=10$ is shown in \fig{histo}. 
The distribution exhibits an asymmetric shape, with a most likely value at $\delta I(\bm{r}_R)\delta I(\bm{r}_T)=0$. 
Due to this asymmetric shape, the distribution cannot be simply characterized by its first moments. In particular,
for $b=7$ and $k_0\ell=10$, we find that the probability to have transmitted and reflected intensity fluctuations
with opposite signs $P(\delta I(\bm{r}_R)\delta I(\bm{r}_T)<0)=0.47$, while we could have
expected a much larger value (above $0.5$) from a naive argument based on the negative sign of the correlation
$C(\bm{r}_R,\bm{r}_T)$. Nevertheless, it is interesting to study statistics under some constraints. 
In particular, we have studied the probability of having $\delta I(\bm{r}_T)<0$ (a spot in the transmitted speckle
darker than the average intensity) under the assumption that $\delta I(\bm{r}_R) > n \bra I(\bm{r}_R)\ket$, with $n\in\left\{0,1,2\right\}$
(\ie for a coinciding spot in the reflected speckle with increasing brightness).
The results are summarized in \tab{proba}. Interestingly, we find that the probability $P(\delta I(\bm{r}_T)<0)$ increases 
substantially with the brightness of the reflected speckle spot, an information that is not contained in the
intensity correlation function. In consequence, if the reflected intensity in a speckle spot is large compared to the average
reflected intensity, the transmitted intensity in the coinciding spot in the transmitted speckle is smaller than 
the average transmitted intensity with a large probability. This result may have implications in the context of light focusing 
through opaque scattering media by wavefront shaping. Indeed, maximizing the intensity in a reflected speckle spot might, 
with a high probability, leads to a minimization of the intensity in the corresponding transmitted spot. A precise study 
with optimized wavefronts (beyond plane-wave illumination) is left for future work.

\begin{table}[!htbf]
   \centering
   \begin{tabular}{|c|c|c|c|c|c|}
   \hline
      \multirow{2}{*}{$b$} & \multirow{2}{*}{$p$} & \multirow{2}{*}{$q$} & \multicolumn{3}{|c|}{$q$}
   \\\cline{4-6}
      & & & $\delta I(\bm{r}_R)>0$ & $\delta I(\bm{r}_R)>\bra I(\bm{r}_R)\ket$ & $\delta I(\bm{r}_R)>2\bra I(\bm{r}_R)\ket$
   \\\hline\hline
      $0.5$ & $0.48$ & $0.57$ & $0.57$ & $0.62$ & $0.67$
   \\
      $1$ & $0.50$ & $0.61$ & $0.64$ & $0.69$ & $0.74$
   \\
      $2$ & $0.47$ & $0.64$ & $0.64$ & $0.67$ & $0.69$
   \\
      $4$ & $0.47$ & $0.65$ & $0.65$ & $0.67$ & $0.67$
   \\
      $7$ & $0.47$ & $0.66$ & $0.66$ & $0.67$ & $0.67$
   \\\hline
   \end{tabular}
   \caption{Probability of having $\delta I(\bm{r}_R)\delta I(\bm{r}_T)<0$ ($p$), of having $\delta I(\bm{r}_T)<0$ ($q$)
   at different optical thicknesses ($b$) and under some constraints ($\delta I(\bm{r}_R)>n\bra I(\bm{r}_R)\ket$ with
   $n\in\{0,1,2\}$) for $\Delta r=0$ and $k_0\ell=10$.}
   \label{proba}
\end{table}

\section{Conclusion}\label{conclusion}

In summary, we have studied analytically and numerically the spatial correlation between intensities measured in
the reflected and transmitted speckles generated by a slab of disordered scattering medium. We have demonstrated the existence of a
reflection/transmission correlation. At large optical thicknesses, the spatial correlation persists and is dominated by
the $C_2$ contribution, thus exibiting a long-range behavior. Interestingly, this correlation takes negative values.  At
small optical thicknesses, the correlation is dominated by a $C_1$-type contribution, which contains the usual $C_1$
term and an additionnal term $C_1'$ involving the ballistic intensity.

The statistical connection between transmitted and reflected speckles might be of interest for
wavefront shaping methods used to focus and image through scattering media. Since for practical implementations
only the reflected speckle can be measured and controlled, a knowledge of the probability to get a bright (dark) spot in 
the transmitted speckle in coincidence with a dark (bright) spot in the optimized reflected speckle could be a great advantage.
As a first step towards this goal, we have studied the statistical distributions of reflected and transmitted intensities,
and have identified situations in which the probability of coincidence of bright and dark spots on opposite sides of the medium is high.

Finally, a refinement of the analytical model would be beneficial to deal with optical thicknesses for which the diffusion approximation fails
to give quantitative results. One possibility could be to developp a semi-analytical approach (coupling analytical expressions and numerical
calculations) based on the radiative transfer equation~\cite{CHANDRASEKHAR-1950} that described accuratly short and long scattering 
paths~\cite{ELALOUFI-2004}.

\section{Acknowledgments}

This work has been initiated by stimulating discussions with Demetri Psaltis and Ye Pu.
We acknowledge Arthur Goetschy for illuminating inputs. The research was supported by LABEX WIFI (Laboratory of
Excellence within the French Program ``Investments for the Future'') under references ANR-10-LABX-24 and
ANR-10-IDEX-0001-02 PSL*. N.F. acknowledges financial support from the French ``Direction G\'en\'erale de l'Armement''
(DGA). 

\appendix

\section{Analytical calculation of $B$}\label{C_1_calculation}

In the reflexion/transmission configuration the usual approximation used to calculate $C_1$ breaks. Indeed in
transmission/transmission or in reflexion/reflexion one usually manages to separate in \eq{C_1_B} integrals over
$\bm{r}_2$ and $\bm{r}_1$ because of the small distance between the two points where we compute the correlation.  In our
configuration, because of the large distance between these two points, we have to calculate explicitly
$B(\bm{r}_R,\bm{r}_T)$.

Using the symmetry of the problem, taking the Fourier transform of the two Green functions over their transverse
coordinates, we simplify \eq{C_1_B} the following way:
\begin{multline}\label{Bsimplified}
   B(\bm{r}_R,\bm{r}_T)=\int \gamma \bra I(z_2) \ket \bra G(z_2,\bm{K}) \ket \bra G^*(z_2-L,\bm{K}) \ket
\\
      \times\exp[-i\bm{K}\cdot\bm{\Delta r}] \ud z_2 \frac{\ud \bm{K}}{(2\pi)^{d-1}}
\end{multline}
where $z_2$ is the depth at which the two fields separate.
To calculate the expression of the Green function in this mixed domain, one can see that:
\begin{align}\nonumber
   \bra G(z_2,\bm{K}) \ket & = \int \bra G(k,\bm{K}) \ket \exp[ikz_2]\frac{\ud k}{2\pi}
\\\label{GreenTF}
     &  = \int\frac{\exp[ikz_2]}{k^2+K^2-k_{\text{eff}}^2}\frac{\ud k}{2\pi}.
\end{align}
\Eq {GreenTF} can be calculated using the residue theorem using the fact that $k_{\text{eff}}^2=k_0^2+ik_0/\ell$ and noting that:
\begin{equation}\label{Residu1}
   \frac{1}{k^2+K^2-k_{\text{eff}}^2}=\frac{1}{(k-k_+)(k+k_+)}
\end{equation} 
where
\begin{equation}\label{kp_kpp}
   k_+=\sqrt{k_0^2+ik_0/\ell-K^2}=k'+ik''
\end{equation}
is the pole with a positive imaginary part (\ie $k''>0$). Thus \eq{GreenTF} can be calculated with the residue theorem
by integrating over a contour composed of the real axis and a upper-half-circle the radius of which  tends to infinity.
We obtain that
\begin{equation}\label{GreenTF2}
   \bra G(z_2,\bm{K}) \ket= \frac{i}{2k_+}\exp(ik_+z_2).
\end{equation}

Following the same steps we can show that
\begin{equation}\label{ConjGreen}
   \bra  G^*(z_2-L,\bm{K})\ket=\frac{-i}{2k_+^*} \exp[ik^*_+(z_2-L)]
\end{equation}
leading to
\begin{multline}\label{Bfinal}
   B(\bm{r}_R,\bm{r}_T)=\gamma\int\bra
      I(z_2)\ket\frac{\exp[2ik'z_2]}{4(k'^2+k''^2)}\exp[-ik^*_+L]
\\
      \times\exp[-i\bm{K}\cdot\bm{\Delta r}]\ud z_2 \frac{\ud \bm{K}}{(2\pi)^{d-1}}.
\end{multline}

Using the expression of the average intensity we have to perform three integrations denoted respectively by $M_1$, $M_2$ and $M_3$:
\begin{align}\nonumber
   M_1 & =\int_0^L(L+z_0-z_2)\exp[2ik'z_2]\ud z_2
\\\label{M_1}
       & = \frac{1 + 2ik'(L+z_0)-\exp[2ik'L](1+2ik'z_0)}{4k'^2},
\\\nonumber
   M_2 & =\int_0^L(z_0+z_2)\exp[2ik'z_2]\ud z_2
\\\label{M_2}
       & = \frac{2ik'z_0-1+\exp[2ik'L][1-2ik'(L+z_0)]}{4k'^2},
\\\nonumber
   M_3 & =\int_0^L \exp(-\frac{z_2}{\ell})\exp[2ik'z_2]\ud z_2
\\\label{M_3}
       & = \ell \frac{1-\exp[2ik'L]\exp[-L/\ell]}{1-2ik'\ell}.
\end{align}
With these expressions, we can rewrite the final expression of $B(\bm{r}_R,\bm{r}_T)$ as a Fourier transform given by
\eq{C_1_B_3}.


\end{document}